\documentstyle[12pt]{article}
\begin{document}
\date{April, 1995}

\title{  Moduli Spaces  in the Four-Dimensional   Topological Half-Flat Gravity
}
\author{
{\bf Mitsuko Abe }
        \\ \\ \\
     {\it Department of Physics, Tokyo Institute of Technology} \\
     {\it Oh-okayama, Meguro-ku, Tokyo 152, Japan}}
\maketitle
\vskip 2.0cm
\begin{center}
{\bf ABSTRACT}
\end{center}
\abstract{
We classify the  moduli spaces of the
four-dimensional topological  half-flat gravity models
by using the canonical bundle.
For a $K3$-surface or $T^4$, they describe an
equivalent class of  a  trio of the  Einstein-K\"ahler forms
( the hyperk\"ahler  forms ).
We   calculate the dimensions  of these moduli spaces
by  using the Atiyah-Singer Index theorem.
We  mention the partition function and the possibility of the observables
in the   Witten-type topological half-flat  gravity model case.
\thispagestyle{empty}
\newpage
{\bf I. Introduction}
\par
Recently, Witten gave  some gravitational versions of topological
quantum field theories~ \cite{witten3}.  These theories are
 important as the effective theories of the ordinal gravity theories.
 For example, he pointed out the  relation  between
the two-dimensional topological gravity models   and the string
theory~ \cite{witten3}. He also conjectured that $N=4$ topological
twisted supersymmetric Yang-Mills theory on four-dimensional
manifold satifies  S-duality and has a link with the  bosonic string or
two-dimensional  rational conformal field theories ~ \cite{witten4}.
They seem to give the new light on the non-perturbative effect of the
string theories and the gravity theories.
\par
Since the work of Witten, there have been several attempts to
construct  four-dimensional topological gravity theories  over
 different kind of the gravitational moduli
spaces~~ \cite{perry}-\cite{kunitomo}.
\par
There  two types of models have been proposed
 for the  four-dimensional half-flat 2-form topological gravity.
(A) Witten-type topological gravity model, which was given by Kunitomo~~
\cite{kunitomo} and
(B) Schwarz-type topological gravity model which we proposed~~
\cite{lee,abe}.
The base of their formalism are given by ref. \cite{capovilla,samuel}.
The interesting relation  between the half-flat gravity
and 2-dim. conformal field theory is investigated by Park~~\cite{park}.
In the previous paper we showed that
by taking the suitable gauge fixing condition and the limit of the
coupling constant  for (B),  the bosonic part of the    moduli spaces of
(B) coincides with that of (A).
These moduli spaces  are  those   of  the Einstein K\"ahlerian
manifolds with  vanishing   real first Chern class.
\par
The purpose  of  this letter is to
calculate the dimensions of the moduli spaces by using
Atiyah-Singer index theorem  and discuss about the possibility of
the topological invariants such as the partition function and
observables. We concentrate our attention for  $K3$-surface and $T^4$ cases
mainly in this paper. Their moduli spaces are identified with the
deformation of a trio of the Einstein-K\"ahler  forms
( the hyperk\"ahler forms )  which  is
related to the Plebansky's heavenly equations~~\cite{plebanski}.
We also discuss   the partition function for the Witten-type model case
and  mension the  the possibility of the observables.
\par
The extension of the algebraic curve with Einstern metric to the
four dimensional case may be the algebraic surfaces with Einstein metrics.
$T^4$ and $K3$ surface belong to the algebraic surfaces and we regard
these models as the simple examples which treat the algebraic surfaces.
\par
As  another aspect,
there have been discovered rich kind of non-compact gravitational
instantons (i.e. ALE ~~\cite{eguchi} or ALF ~~\cite{hawking})
which satisfy these equations.
In this paper, we will treat  the compact manifolds  only.
In the near future, we will extend our investigation to non-compact case.
\newpage
{\bf  II. The Four-Dimensional Half-Flat Topological Gravity Models}
\par
We derive the moduli spaces of the four-dimensional
half-flat topological gravity models,
which  are  partially contained in ref.\cite{abe}.
There are two types of models : (A) Witten-type action given
by Kunitomo~~\cite{kunitomo} and (B) Schwarz-type action  proposed
by us \cite{lee,abe,capovilla,horowitz}.
We follow the notations of  ref. \cite{kunitomo,abe}.
Fundamental fields in these models are  a trio of $su(2)$ valued 2-form
$\Sigma^k=\Sigma^k_{\mu\nu} dx^\mu  \wedge dx^\nu$ and a trio of $su(2)$
valued 1-form $\omega^k=\omega^k_\mu dx^\mu$.
The curvature tensor  $F^k= F^k_{\mu\nu} dx^\mu
\wedge  dx^\nu \equiv
d\omega^k+ (\omega \times \omega)^k = d\omega^k + f^{ijk}
\omega^i \wedge \omega^j$
($f^{ijk}$ is a structure constant of $SU(2)$ and it will be $\epsilon^{ijk}$
in later).
Varying  the    actions of (A) or (B)   with respect to
$\omega^k $ and $\Sigma^k$  fields,
we obtain the following  equations\cite{kunitomo,abe};
\begin{equation}
F^k=0,~~D\Sigma^k=0\ ,
                                        \label{eq:ins2}
\end{equation}
\begin{equation}
    {}^{t.f.} \Sigma^i \wedge \Sigma^j
    \equiv \Sigma^{(i} \wedge \Sigma^{j)}
    - {1 \over 3} \delta_{ij} \Sigma^k \wedge \Sigma_k = 0,
\label{eq:eight}
\end{equation}
\par
$M_4 $  is supposed to be a four-dimensional compact
K\"ahler manifold with its real  first Chern class $c_1(M_4)_R= 0$,
which is the sufficient condition for the existence
of the moduli spaces \cite{abe}.
{}From the Bogomolov decomposition theorem~\cite{bogomolov},
 complex tori $T^4$~ or $K3$ are   possible as the covering space of $M_4$.
The field $\omega^k= \omega^k_\mu dx^\mu$ is restricted to
a self-dual chiral part of a frame connection ${}^{(+)}\omega^{a b}_\mu$~
 i.e.,we consider only the principal bundle of  oriented orthonormal frames
$P_{SO(4)}$  over $M_4$  with structure group $SO(4)$.
{}From the assumption for $M_4$, at least the reduction of
the structure group $G=SO(4) \rightarrow
U(2)$ and  of the bundle  $P_{SO(4)} \rightarrow
P_{U(2)} $
 are possible \cite{kobayashi}.
Thus,   only  $U(1)$ component of $\omega^k$ exists as the self-dual part.
\begin{equation}
{}^{(+)}\omega^{ab}=\bar \eta_k^{ab} \omega^k,~
\omega^k \in su(2)  \times \wedge^1
\rightarrow
\omega^1 \in u(1)  \times \wedge^1,~ \omega^2= \omega^3=0,
\end{equation}
where $\bar \eta^i_{ab}$ is the t'Hooft's $\eta$-symbol
\cite{t'hooft}.
\par
Similarly, we get the following reductions
\footnote{In the previous paper~\cite{abe}, we are setting $\Sigma^k$
to be represented by $su(2) \otimes \Lambda^2$. It will be more convenient to
use the reduction forms $1 \times \Lambda^2$ or $u(1) \otimes \Lambda^2$ for
them to see the dimensions of the each cohomology group which will appear in
the later.} for $\Sigma^k$, which supposed to be self-dual part of $SO(4)$ :
\begin{equation}
\Sigma^i \in su(2)  \otimes \Lambda^2
\rightarrow
\Sigma^1 \in 1  \otimes \Lambda^2,~
\Sigma^2~{\rm and}~ \Sigma^3 \in u(1)  \otimes \Lambda^2.
\end{equation}
{}From eq. (2),  $\Sigma^k$
 comes from a   vierbein  $e^a= e^a_\mu dx^\mu$~\cite{capovilla} :
\begin{equation}
 \Sigma^k(e)  = - \bar \eta^k_{ab} e^a \wedge e^b \propto
g_{\alpha \bar \beta}  J^{k \bar \beta }_{\bar \gamma} dz
^\alpha \wedge d\bar z^ {\bar  \gamma}.
\label{eq:nine}
\end{equation}
$\{ J^k \}$ represents a complex structure   or two  almost complex structures
which  satisfy the  quaternionic relations and $g_{\alpha \bar \beta}$ is a
hermite  symmetric metric.
The self-dual part of the Riemannian
tensors  is related  to the curvature tensor  $F^k$
\cite{kunitomo,abe}.
It becomes  zero (i.e.,Riemannian  half-flat) when it satisfies the eq. (1) and
(2)
\cite{kunitomo,abe}~; ~
\begin{equation}
{}^{(+)} R_{\mu \nu \rho \tau}= 0.
\end{equation}
{}From the Hitchin's theorem~~\cite{besse},
 $(M_4, g)$ is covered by a flat  4-torus
or K\"ahler-Einstein K3-surface  or
K\"ahler-Einstein  K3-surface /$Z_2 $ or
K\"ahler-Einstein K3-surface /$Z_2 \times Z_2$.
\par
Before we proceed, it will be useful to introduce
general spin bundles $\Omega^{m,n}$ \cite{besse}
and the canonical bundle $K$\cite{hirzebruch}.
A U(1) part deformation of    the frame connection is denoted by
$\omega^1_f$. It  belongs to   $K^\ast \otimes  \Lambda^1$.
\begin{equation}
 \omega^1_f \in \Omega^{0,2}\otimes \Lambda^1
 \rightarrow  \omega^1_f \in  P_{U(1)} \times Ad_{u(1)} \otimes \Lambda^1
 \cong K^\ast  \otimes \Lambda^1,
\end {equation}
which comes from the following isomorphism \cite{kobayashi2}:
\begin{equation}
P_{SO(4)} \times Ad_{so(4)}  \cong  M_4 \times so(4) / Adj
\equiv  M_4 \times \wedge^2 R^4 / Adj
\cong \wedge^2 TM_4^\ast \cong \Omega^{2,0} \oplus \Omega^{0,2}
\end{equation}
Similarly, $\{\Sigma^k_f\}$ the deformations of $\Sigma^k$  are
  represented by   $\Sigma^1_f \in {\bf R}S \otimes \Lambda^2, \Sigma^2_f
{}~{\rm and}~  \Sigma^3_f \in K^\ast \otimes \Lambda^2$, where   $S$ is
a certain parallel section of $\Omega^{0,2}$ ~~\cite{ito}.
\begin{equation}
\Sigma^1_f, \Sigma^2_f, \Sigma^3_f \in
 \Omega^{0,2} \oplus \Lambda^2
\rightarrow  \Sigma^1_f, \Sigma^2_f, \Sigma^3_f \in
({\bf R}S \oplus 2K)^\ast \otimes \Lambda^2.
\end{equation}
The moduli spaces of  (A) coincide with  those of the bosonic part of (B)
 and are classified into two cases by using the property of the
canonical bundle $K$ \cite{abe}.
\\
\\
 Case(1)  when the canonical  bundles  are trivial:
$M_4$ is a  $K3$-surface  or $T^4$.
\\
On these two manifolds, the  reductions of $P_{U(2)}$
 are possible  due to the fact that $K$ and $P_{U(1)}$
  are trivial when they have  Einstein-Kahler metrics.
 These manifolds  are called hyper-K\"ahlerian.
\par
The  moduli space is the equivalent class of  a trio of the
Einstein-K\"ahler forms  (the hyperk\"ahler forms) $\{\Sigma^k(e)\}$~
\cite{abe}~:~~
\begin{equation}
{  \cal M}(\Sigma)
=\{ \Sigma^k
\mid \Sigma^1 \in 1 \otimes \wedge^2~,
\Sigma^2, \Sigma^3  \in u(1) \otimes \wedge^2~,
 {}^{t.f.} \Sigma^i \wedge \Sigma^j=0,
 d \Sigma^k=0 \} / diffeo.
\end{equation}
\\
Case(2) when the  canonical bundles are not trivial:
 $M_4$ is  $K3/Z_2$, or   $ K3/Z_2 \times Z_2 $,
or  $T^4  / \Gamma$ ($\Gamma $ is some discrete group).
\par
As   $K$ is not trivial on
these manifolds, the reduction of $P_{U(2)} \rightarrow P_{{SU(2)}_R}$
is not possible.
These manifolds are called as the locally hyperk\"ahlerian.
Their moduli spaces are given by
\begin{eqnarray}
{\cal M}(\Sigma,~\omega )
&=&\{ \Sigma^k,~\omega^1
\mid \omega^1 \in u(1) \otimes \Lambda^1, F^1_ {\mu\nu} =0, ~~
\Sigma^1 \in 1 \otimes \wedge^2~,
\Sigma^2, \Sigma^3  \in u(1) \otimes \wedge^2~,
\cr
&~&~~~~~~~~{}^{t.f.} \Sigma^i \wedge \Sigma^j=0,
{}~d \Sigma^1=0, D\Sigma^2=D\Sigma^3=0 \} / diffeo. \times U(1).
\end{eqnarray}
These moduli spaces  have  the  bundle structure
such  as the base is $ {\cal M} (\omega^k)$
with a fibre $ {\cal M} (\Sigma)$.
\begin{eqnarray}
 { \cal M}(\Sigma)=\{ \Sigma^k
&\mid& \Sigma^1 \in 1 \otimes \wedge^2~,
\Sigma^2, \Sigma^3  \in u(1) \otimes \wedge^2~,
\\ \nonumber
&~&~~~~~~ {}^{t.f.} \Sigma^i \wedge \Sigma^j=0,
 d \Sigma^1=0,  D \Sigma^2=D \Sigma^3=0  \} / diffeo.,
\\ \nonumber
 { \cal M}(\omega) =\{ \omega^1
&\mid& \omega^1 \in 1 \otimes \Lambda^1,
{}~F^1_{\mu \nu}=0  \} / U(1).
\end{eqnarray}
{\bf III. BRST quantization    in the Witten-type model on $T^4$ or $K3$ }
\par
We discuss about the partition function of the Witten-type
model on $K3$ or $T^4$,  whose BRST symmetry and
the action   is  given by \cite{kunitomo}.
(We will report   the partition function for the Schwarz-type model
 in the next paper \cite{mmabe}. )
The action of the Witten-type model
reduces to
\begin{eqnarray}
S_0^{\rm red}
&=& \int_{M_4} ( \pi_i   \wedge d\Sigma^i+ {1 \over 2}
\pi_{ij}  {}^{ t. f.} \Sigma^i \wedge \Sigma^j
- \chi_i \wedge  d\Psi^i- {1 \over 2}\chi_{ij}  {}^{t.f.}
\Sigma^i \wedge \Psi^j )
\cr
&=&
\{ \int_{M_4} ( \pi_i \oplus \pi_{ij} ) \wedge  D_1 \Sigma^j
- (\chi_i  \oplus \chi_{ij})\wedge  D_1 \Psi^j \},
\end{eqnarray}
 where  $ D_1  \Sigma^j_f \equiv
(\hat D_1 \oplus \tilde D_1^{i} ) \Sigma^{j}_f
 \equiv  ( d \oplus {}^{t.f.} \Sigma_0^{i} \wedge )   \Sigma^{j}$.
This action is invariant under the diffeomorphism  transformations and
the  supersymmetry.  These transformations are invariant under
 the redundant diffeomorphism  transformations of them.
The symmetries of this model is interpretable as
${diffeo \otimes super. \over  red.~ diffeo.}$.
\par
We use the  decomposition $\Sigma^k = \Sigma^k_0 + \Sigma^k_f$
of the $\Sigma^k$ field
to calculate of the partition function
where $\Sigma^k_0$ is a back ground solution of the
 equations of motions.
Gauge fixing conditions which we set  are $  D_0 ^\ast  \Sigma^k =0$
for the diffeomorphism
and   $  D_0 ^\ast \Psi^k =0$ for the redundant
diffeomorphism and  $  D_1   \Sigma^k =0$
for the super symmetry (except for the redundunt diffeo symmetry).
$D_0$  is defined by $  D_0  c^{\mu} \equiv { {\cal L}}_c  \Sigma^k$.
 $\ast$  denotes the Hodge star dual operation and
${\cal O}^\ast \equiv - \ast {\cal O} \ast $  is
the adjoint operator of ${\cal O}$. The dimensional countings of the
fundamental fields are given  in the Table 1.
In this model, there are  the  redundant symmetries  such as
$
\delta\ast\pi_k=D_2^\ast\lambda_k$~ and  $\delta\ast\chi_k=D_2^\ast\eta_k
$
due to
$
D_2  \hat D_1 \Sigma^k \equiv d^2   \Sigma^k =0 $~~
 and  $D_2  \hat D_1 \Psi_k= d^2\Psi^k= 0
$ \cite{kunitomo}.  $\lambda_i$ and  $ \bar \lambda^i $
($\eta_i $ and $\bar \eta^i$)
  are a ghost and an anti-ghost
of these redundant symmetries
  with $( 2 K\oplus {\bf R}S)^\ast \otimes \Lambda^4 $ representations.
To  fix these symmetries, we set
$
D_2 \ast \pi_k=0~
{\rm and}~
D_2  \ast \chi_k=0.
$
Thus the quantum action  $S_q$ is given by
\begin{equation}
S_q = S_0^{\rm red}+ S_{\rm g. f.},
\end{equation}
\begin{equation}
S_{\rm g. f.}
= \int_{M_4} \delta_B \{ \ast b \wedge  D_0^\ast  \Sigma^k_f
                       + \ast\bar  \lambda^k \wedge  D_2 \ast \pi_k
                       - \ast \beta \wedge  D_0^\ast \Psi^k
                       -  \ast \bar \eta^k  \wedge  D_2 \ast \chi_k \} .
\end{equation}
$\delta_B$ denotes the BRST-transformation.
We are now ready to evaluate the partition function :
\begin{equation}
Z= \int  {\cal D}X \exp(-S_q),
\label{eq:fseven}
\end{equation}
where ${\cal D}X$ represents the path integral over the fields
$ \Sigma^k_f$, ghosts, anti-ghosts and N-L fields.
In general, these fields contains zero modes and non-zero modes.
We introduce the following  the deformation complex
\footnote{We thank  T. Ueno for pointing out
  the possibility of using the deformation complex
and the Index  theorem in these models\cite{ueno}.
Before us, he composed the similar deformation complex and
calculated the index for these models.}.
The zero modes   are the elements of
the cohomology groups of this complex.
\begin{eqnarray}
&V_0 & = \Omega^{1, 1},~~
V_1 = ( 2 K\oplus {\bf R}S)^\ast  \otimes \Lambda^2 ,
\\ \nonumber
&V_2 &= (2 K  \oplus 2 K ^{\otimes 2}  \oplus {\bf R}S)^\ast
\otimes \Lambda^4
\oplus (2K  \oplus {\bf R}S)^\ast  \otimes \Lambda^3,~~
V_3 =  0 \oplus (2 K \oplus {\bf R}S)^\ast \otimes \Lambda^4.
\end{eqnarray}
\begin{equation}
  0   \stackrel{D_{-1}} \to
C^\infty ( V_0 )
   \stackrel{D_0} \to
C^\infty (V_1)
    \stackrel{D_1} \to
C^\infty (V_2)
    \stackrel{D_2} \to
C^\infty (V_3)
   \stackrel{D_3} \to  0  .
               \label{eq:fifive}
\end{equation}
$D_{-1}$ and $D_{3}$ are identically zero
operators.
We can easily check the ellipticity of the deformation complex.
We may then define the cohomology group.
\begin{equation}
 H^i \equiv {\rm Ker}\, D_i/{\rm Im}\, D_{i-1}= {\rm Ker}\triangle_i,
                                            \label{eq:fiseven}
\end{equation}
where $\triangle_i = D_{i-1}^{} D_{i-1}^* + D_i^* D_i^{}$
and  $D_1^\ast D_1 \cong \hat D_1^\ast \hat  D_1 +
\tilde D_1^\ast \tilde  D_1$ .
 The dimensions  of $H^i$   are finite and represented by $h^i$.
 $H^1$ is exactly identical with  ${\cal T( M }(\Sigma) )$
the tangent space  of ${\cal M}(\Sigma)$  :
\begin{equation}
  T({\cal M}(\Sigma))
  = \{  \Sigma^k_f \vert \Sigma^k_f \in (2 K \oplus {\bf R}S)^\ast  \otimes
\wedge^2,  D_1   \Sigma^k_f = 0 \} /
              diffeo.   \ .
                                                  \label{eq:fitwo}
\end{equation}
\par
When expanded out   by using the properties of $\delta_B$
\cite{kunitomo},   the quantum action is given  by
\begin{eqnarray}
S_q & =&
 \ast B \wedge    T B^t -  \ast F \wedge    T F^t
\cr
&+& \ast \beta \wedge  \Delta_0 \gamma-  \ast b \wedge   \Delta_0 c
\cr
&+&  \ast \bar \eta^k  \wedge  \Delta _3 \eta_k -
 \ast  \bar \lambda^k  \wedge  \Delta_3 \lambda_k +
{\rm other~ higher~ order~ terms},
\end{eqnarray}
with some field redefinitions.
\[  T=\left(
\begin{array}{@{\,}cccc@{\,}}
0    &- D_0^\ast        & 0                   & 0         \\
-D_0 & 0                &  -(\hat D_1^\ast \oplus  \tilde D_1^\ast) & 0    \\
0    &-(   \hat D_1\oplus  \tilde D_1)   & 0  & - D_2^ \ast        \\
0    &  0               & -D_2                & 0
\end{array}
\right). \]
We integrate over non-zero modes.
The Gaussian integrals over the commuting
$  \beta -  \gamma$~ and $\bar \lambda^k-\lambda^k $
sets of fields  cancel
with the contribution  coming from
the anti-commuting sets of   $ b-  c $~  and ~
$\bar \eta^k- \eta_k$~ .
We integrate  over the remaining
$ B \equiv (\pi_c,~~ \Sigma^k_f,~~\ast (\pi^k \oplus \pi_{ij}),~~
(\pi_{\lambda})_k ) $
-system  and
$F \equiv \{ \chi_{\gamma},~~\psi^k,~~\ast(\chi_i \oplus\chi_{ij}),~~
(\chi_{\eta})_k  \}$- system
 by taking ${\rm det}T ={\rm det}^{1 \over 2} (T^\ast T)$
and  using the nilpotency $D_iD_{i+1}=0$
\cite{blau}. Their contributions cancelled out each other.
If the dimension of  moduli space is zero,
then  there may exist the non-trivial partition with
  projecting out of the other zero modes.
\begin{equation}
Z
= \Sigma^\prime_{\rm instanton}
{
\{
({\rm det} \Delta_0)
({\rm det} \Delta_1)
({\rm det} \Delta_2)
({\rm det} \Delta_3)
\}
^{1 \over 4}
\over
\{
({\rm det} \Delta_0)
({\rm det} \Delta_1)
({\rm det} \Delta_2)
({\rm det} \Delta_3)
\}
^{1 \over 4}
}
\cdot
{
\{
({\rm det} \Delta_0)
({\rm det} \Delta_3)
\}
\over
\{
({\rm det} \Delta_0)
({\rm det} \Delta_3)
\}
}
 =  \Sigma^\prime_{\rm instanton} \pm 1
\label{eq:fnine}
\end{equation}
\par
Later, however,  we show that
 $h^1$  is non-zero  on $K3$~   and  on $T^4$.
Thus, there arise fermionic zero-modes and bosonic zero-modes
 whose numbers are equal to  $h_1$ in this model.
The path integral over the fermionic( bosonic) zero modes makes
the partition function trivial (divergent).
\par
 Some appropriate combinations
of the following  three ways  can avoid these situations :
\\
1) projecting out i.e.,  gauge fixing of  the global symmetries caused by the
zero modes
\\
2) evaluation of the interaction terms of the quantum action
 which contains the appropriate zero modes
\\
3) introducing some functional ${\cal O}$  called `observable'
and calculating the vacuum  expectation value of it \cite{witten3}.
The path-integration over them
may provide non-trivial information
to distinguish differential
structures on  these  manifolds.
\par
Myers et al. have already proposed some  `observable'
in their  gravity model\cite{myers}.
Their moduli space is the equivalent class of the  Riemannian
half-flat metric  with respect to the diffeomorphism.
Their `observable'  is composed of  the BRST-extension of the Riemannian
 curvature form on $M_4$ and it  might express the topological
invariants made  of  curvatures  on
 the tangent moduli space.
The BRST symmetries of  Myers et. al.'s model   can be
realized by using  the  composite fields   of  $\Sigma^k$~
and  its super partner~ $\Psi^k$ on $K3$ or $T^4$  in Kunitomo's model
since a metric can be made of $\Sigma^k$~\cite{kunitomo}.
Thus the observables  given by Myers et. al are  also
adopted to the Kunitomo's model.
The evaluation of them in the Kunitomo's model
will give some topological information.
\\
{\bf VI. The dimensions of the Moduli Spaces }
\par
Let $K(g)$ be the moduli space of  Einstein-K\"ahler forms,
$\epsilon(g)$ be the moduli space of  Einstein metrics
and   $C(g)$ be the moduli space of complex structures, respectively.
All of them are the equivalent classes  with respect to the  diffeomorphism.
At first, we   quote the result  about the dimension of $K(g)$
 when $M_4$ is  K\"ahlerian manifold
with $c_1(M)_R=0$,  which is   given by ~\cite{besse}.
Then we show the dimension of ${\cal M}(\Sigma)$ or ${\cal M}(\Sigma, \omega)$
 by using the Atiyah-Singer Index theorem
 and clarify the difference between ${\cal M}(\Sigma)$ and ~$K(g)$,
which is partially contained in ref. \cite{abe}.
\par
When $c_1(M)_R=0$,
the deformation of the K\"ahler class
with a fixed complex structure induces a deformation of a Einstein
metric from the Calabi-Yau theorem \cite{calabi}.
 The deformation of  Einstein-K\"ahler forms
$\{ \Sigma \}$  consists of
that  of Einstein metrics $\{ g \} $ and  complex structures
$\{ J \}$ and needs a careful examination of
its degenerated part ;
$
\delta \Sigma = \delta g \circ J + g \circ  \delta J.
$
The dimensions of $K(g)$, $\epsilon(g)$ and $C(g)$  for the
K\"ahlerian manifolds   with  $c_1(M)_R=0$  are given  by
\begin{eqnarray}
{\rm dim.} &C& (g)= 2{\rm dim.}H^1_C(M, \Theta),
\\
{\rm dim.} &\epsilon& (g)= {\rm dim.} H^{1,1}_R (M, {\bf J}) -1 +
                       2{\rm dim.}H^1_C(M, \Theta)-2{\rm dim.} H^{2,0}_C(M,
{\bf J}),
\\
{\rm dim.} &K&(g)= {\rm dim.} H^{1,1}_R (M,{\bf J}) -1 +
2{\rm dim.}H^1_C (M, \Theta).
\end{eqnarray}
where $\Theta={\cal O}(TM_z)$
i.e. the sheaf of the germs of holomorphic vector fields.
  $TM_C=TM_4 \otimes C=TM_z \oplus \bar {TM_z}$ ~
 with  formal splittings \cite{shanahan}~:~$~TM_z=  L_1+ L_2,
\bar {TM_z}=\bar L_1+ \bar L_2$. $L$ and $\bar L$ are the line bundles.
\par
We show how to calculate the dimension of the moduli space of
 ${\cal M}(\Sigma)$.  The calculation of the index for  ${\cal M}(\Sigma)$
is  common to both of case(1) and case (2)
because the index is independent on the connections.
By applying the Atiyah-Singer index theorem\cite{shanahan} ,
we obtain
\begin{eqnarray}
 {\rm Index}&=&  \Sigma^3_{i=0} (-1)^i h^i
\\ \nonumber
&=&\int_{M_4} \frac{ {\rm td} (TM_4 \otimes {\bf C} ) }
      {{\rm e} (TM_4) } \cdot  {\rm ch} \{ \sum_{n=0}^3 \oplus (-1)^n V_n \}
\\ \nonumber
&=& 2 \chi + 7 \tau  \rightarrow 2 \chi - 7 \mid \tau \mid  \, ,
                                             \label{eq:finine}
\end{eqnarray}
where ${\rm ch},$ ${\rm e}$ and ${\rm td}$ are the Chern
character, Euler class and Todd class of the various vector bundles
involved.
The index is determined by the Euler number $\chi=\int_{M_4}
x_1 x_2$ and  Hirzebruch signature $\tau=\int_{M_4}{x_1^2+x_2^2 \over 3} $.
$x_i$ denotes the first Chern classes of $L_i$ or $\bar {L_i}$
By changing $\tau \rightarrow \mid \tau \mid $, this index can be
also adopted to manifolds with the opposite orientation.
In the above calculation, we use the character of the general spin
bundles \cite{romer} and that of a canonical bundle :
$
{\rm ch}(\Omega^{0,l})=A_l+2B_l(y_+)^2+{2 \over 3}C_l(y_+)^4 + \cdots,
$
with $A_l=l+1,~B_l=\sum_{k=0}^l(k-{1 \over 2}l)^2,~
C_l=\sum^l_{k=0}(k-{1 \over 2}l)^4$.
 $y_+$ is replaced by $y_-$ for ${\rm ch}(\Omega^{l,0})$.
 $y_\pm$ is given  in terms of $x_i$  due to the relation
between  the  fundamental spinor representation and the adjoint
representation of the Unitary group~\cite{shanahan}:
$
{\rm ch} (L_i )=\exp {x_i},~
x_i=c_1(L_i)= -c_1( \bar L_i)~ {\rm and}~  y_{\pm}={1 \over 2}(x_1 \pm x_2).
$~~
$
{\rm ch}(K^\ast)= {\rm ch}(L_1^\ast \otimes L_2^\ast)=\exp \{ -(x_1+x_2)\}.
$
In terms of $x_i$,
$
td(TM_c)=1-{x_1^2+x_2^2 \over 12}+{\rm higher~ order ~terms},
{}~~~
e(TM_4)\mid_{M_4}=x_1 x_2 .
$
It is easy to see  that
$H^0$ is equivalent to the space of the Killing vectors (Isometry),
$ H^2=2H^2(M_4,{\cal O} (K^\ast))+H^2(M_4, R)$
and $H^3 =2H^4(M_4,{\cal O}(K^\ast))+H^4(M_4, R)$.
We use the  notation
${\rm dim.}H^i(M_4, {\cal O}(K^\ast)
) =b_i(K^\ast)$
 for the dimension of the cohomology
group of the twisted de Rham complex~:~
\begin{equation}
\cdots \stackrel{D} \to
C^\infty (K^\ast \otimes  \wedge^2  TM^\ast )
    \stackrel{D} \to
C^\infty (K^\ast  \otimes \wedge^3  TM^\ast)
    \stackrel{D} \to
C^\infty (K^\ast \otimes \wedge^4  TM^\ast)
   \stackrel{} \to  0  .
               \end{equation}
\begin{equation}
{\rm dim.} {\cal M}(\Sigma)= h^1 = -2 \chi + 7 \mid \tau \mid + {\rm dim.~of
{}~the~
Isometry}
+2b_3(K^\ast)+b_3  - 2b_4(K^\ast)-b_4  \ .
                                             \label{eq:fiten}
\end{equation}
To  know  $b_i(K^\ast)$,  we use
the following equations and $b_{4-i}(K^\ast) \cong  b_i(K)$.
\begin{eqnarray}
(A)b_1(K^\ast)&=&{\rm dim.} H_{\bar \partial}^0 (M_4,{\cal O}
( TM^\ast_z \otimes \wedge^2  TM^\ast_z))
+{\rm dim.} H^1_{\bar \partial} (M_4, {\cal O}(  \wedge^2  TM^\ast_z))
\\ \nonumber
&=& {\rm dim.} H_{\bar \partial}^2 (M_4, \Theta) + b_{1,2},
\\ \nonumber
(B)b_2(K^\ast) &=&{\rm dim.} H_{\bar \partial}^2 (M_4,{\cal O}
(  \wedge^2  TM^\ast_z))
+{\rm dim.} H_{\bar \partial}^0 (M_4, {\cal O}(\wedge^2
TM^\ast_z \otimes  \wedge^2  TM^\ast_z))
\\ \nonumber
&+&{\rm dim.} H_{\bar \partial}^1 (M_4,{\cal O}
(TM^\ast_z \otimes  \wedge^2  TM^\ast_z))
\\ \nonumber
&=&{\rm dim.} H_{\bar \partial}^{2,1} (M_4,{\cal O}(   TM^\ast_z))
+
{\rm dim.} H_{\bar \partial}^0 (M_4, {\cal O}(\wedge^2
TM^\ast_z \otimes  \wedge^2  TM^\ast_z))
  + b_{2,2}
\\ \nonumber
&=&H_{\bar \partial}^1(M_4, \Theta)+ p^2+b_0,
\\ \nonumber
(C)b_3(K^\ast) &=&{\rm dim.} H_{\bar \partial}^1 (M_4, {\cal O}( \wedge^2
TM^\ast_z \otimes
\wedge^2  TM^\ast_z))
\\
\nonumber
&+& {\rm dim.} H_{\bar \partial}^2 (M_4,  {\cal O}(TM^\ast_z \otimes
\wedge^2  TM^\ast_z))
\\ \nonumber
&=&   {\rm dim.} H_{\bar \partial}^0(M_4,\Theta)
+ ({7 \over 2}\tau+{3 \over 2}\chi)+{\rm dim.}
H_{\bar \partial}^1(M_4, \Theta)-b_{1,2}
\\ \nonumber
&=&   {\rm dim.} H_{\bar \partial}^0(M_4,\Theta)
+ (-{7 \over 2}\mid \tau \mid +{3 \over 2}\chi)+{\rm dim.}
H_{\bar \partial}^1(M_4, \Theta)-b_{1,2},
\\  \nonumber
(D)b_4(K^\ast)& =&{\rm Index~ of~ eq.(27)}
-\sum_{i=0}^3(-1)^ib_i(K^\ast),~(E)~ b_0(K^\ast)=b_{0,2}.
\end{eqnarray}
$p^2$ is the 2-th plurigenus~\cite{hirzebruch}.
 They are derived by the Serre  duality \cite{hirzebruch} and
the  index theorem   applied to the twisted Dolbeault complex\cite{hirzebruch}
or the twisted de Rham complex\cite{shanahan}.
\begin{equation}
{\rm dim.} H_{\bar \partial}^{p,q}(M_4, {\cal O}(F))
\cong
{\rm dim.} H_{\bar \partial}^{2-p,2-q}(M_4, {\cal O}(F^\ast)),
\end{equation}
\begin{equation}
H_{\bar \partial}^{p,q}(M_4,{\cal O}(F)) \equiv
H_{\bar \partial}^{q}(M_4,{\cal O}(F \otimes \wedge^p  TM^\ast_z )),
\end{equation}
\begin{equation}
{\rm dim.} H_{\bar \partial}^{q}(M_4,{\cal O}( \wedge^p  TM^\ast_z ))
= {\rm dim.} H_{\bar \partial}^{p,q}(M_4, C),
\end{equation}
where $F$ is a holomorphic bundle.
We used the result of ref. \cite{ito} to
know the the number of $p^2$ and $b_{2,0}$ for $K3/Z_2 \times Z_2$
(see Table 2).
\par
We introduce the notation $ {\rm dim. } \tilde {\cal M}(\Sigma) \equiv
{\rm dim. } {\cal M}(\Sigma)-1 $   to  remove a scale factor.
For  the case (1),  dim. $ \tilde {\cal M}(\Sigma)$
which we derived  and   dim. $K(g)$, dim. $\epsilon(g)$ and dim. $C(g)$
given by \cite{besse} are summarized in
\begin{center}
\begin{tabular}{|c|c|c|c|c|}  \hline \hline
\it {} & {\rm dim.} $ \tilde {\cal M} (\Sigma) $ &
{\rm dim.} $K(g)$  & {\rm dim.} $\epsilon (G)$ & {\rm dim.} $C(G)$
\\ \hline
\it $K3$ &  60 & 59 & 57 & 40
\\ \hline
\it  $T^4 $     &12 & 11 & 9 & 8
\\ \hline
\end{tabular}
\end{center}
\par
The difference  between  $ \tilde {\cal M}(\Sigma)$  of the case(1)
and  $ K(g)$ is  as follows~;~
$\tilde {\cal M}(\Sigma)$ represents the moduli space of
hyperk\"ahler forms i.e., the definition of $\tilde {\cal M}(\Sigma)$
describes  a set of $(g, J^1, J^2, J^3)$ or  equivalently
$(\Sigma^1, \Sigma^2, \Sigma^3)$, which takes into account the
degrees of freedoms of how  one can choose $g$, a
trio of $g$-orthogonal  complex structures.
\par
On the other hand, $K(g)$ designates $(g, J^1)$ or
equivalently  $(\Sigma^1)$ only.
The degrees of a trio of g-orthogonal complex strucrtures
which satisfy the quaternionic relations for a fixed $g$ is 3.
Namely, for a fixed $g$,
\begin{eqnarray}
 &\{& g-{\rm orthogonal~quaternionic~ almost~ complex~ structures}~J \}
\\ \nonumber &\cong& SO(4)/U(2)
 \cong S^2~ \cong~{\rm Im} {\bf H} \mid_{x_1^2+x_2^2+x_3^2=1},
\end{eqnarray}
\begin{eqnarray}
{\rm Im} {\bf H}
 \equiv
\{&J=\sum_{i=0}^3 x_i  \tilde J^i  &\mid
\tilde J^i \tilde  J^j
= - \tilde J^j \tilde  J^i=J^k~
 (i,j, k ~ {\rm cyclic}), \\ \nonumber &( x_1, x_2, x_3)\in R^3& ,
 (\tilde J^i)^2=(\tilde J^j)^2=(\tilde J^k)^2=-1 ~
  \} .
\end{eqnarray}
For the hyperk\"ahlerian manifold, a trio of almost complex structures
reduces to  a trio of complex structures.
The degrees of freedom of  how one can choose $J^1$ for a fixed $g$
is given by $  {\rm. dim.} S^2 =2$.
The degrees of freedom of how  one can obtain $J^2 $ which is  orthogonal
to $J^1$ for
a fixed pair $(g, J^1)$ is given by $ S^1$ over $S^2 =2$.
$J^3$ is automatically arranged after $(g, J^1, J^2)$
 are fixed.
\par
The moduli space ${\cal M} (\Sigma) $ has a bundle structure
which has     the fiber $(J^1,J^2,J^3)$  over
the base manifold $ \epsilon(g) $ .
\begin{eqnarray}
{\rm dim.} \tilde {\cal M}(\Sigma)&=&{\rm dim.}~K(g)+ {\rm dim.} S^1
\\ \nonumber &=&{\rm dim.}~ \epsilon(g)+ 2{\rm dim.} H^{2,0}_C(M,{\bf J}) +
{\rm dim.} S^1
\\ \nonumber &=&{\rm dim.}~ \epsilon(g)+ {\rm dim.} S^2 + {\rm dim.} S^1 .
\end{eqnarray}
For the case (2), The dimension of the moduli space is given by
\begin{equation}
{\rm dim.} {\cal M}(\Sigma, \omega)
={\rm dim.}{\cal M}(\Sigma) +
{\rm dim.} {\cal M}(\omega),
{}~~{\rm dim.}  {\cal M}(\omega)= b_1(K^\ast).
\end{equation}
${\cal M}(\Sigma)$ describes one K\"ahler form and two
 almost K\"ahler forms which satisfy the quaternionic relations, whose
dimension is given by eq.(28).
${\rm dim.}{\cal M}(\Sigma)$  derived by us  and ${\rm dim}{\cal M}(\omega)$
and  $ {\rm dim.}K(g)$,
${\rm dim.}\epsilon(g)$ given by  \cite{ito,besse} are summarized  in
\begin{center}
\begin{tabular}{|c|c|c|c|c|c|}  \hline \hline
\it {} &{\rm dim.} {\cal M} $(\omega)$ & {\rm dim.} $ \tilde {\cal M} (\Sigma)
$ &
{\rm dim.} $K(g)$  & {\rm dim.} $\epsilon (G)$ & {\rm dim.} $C(G)$
\\ \hline
\it $K3/Z_2$ & 0& 30 & 29 & 29 & 20
\\ \hline
\it  $K3/Z_2 \times Z_2 $     &0 & 15 & 14 & 15 & 10
\\ \hline
\end{tabular}
\end{center}
The difference  between  {\rm dim.} $K(g)$ and {\rm dim.}
$ \tilde {\cal M}(\Sigma)$  corresponds to {\rm dim.} $S^1=1$ as before
but the  difference between ${\rm dim.}K(g)$ and ${\rm dim} \epsilon(g)$
is no longer represented by {\rm dim}.$S^2$
since $ J^1 \not= x_1 \tilde J^1+x_2 \tilde J^2+ x_3 \tilde J^3 $
for a complex structure $J^1$.
\begin{eqnarray}
{\rm dim.}\tilde {\cal M}(\Sigma)&=&{\rm dim.}~K(g)+ {\rm dim. }S^1
\\ \nonumber
&=&{\rm dim.}~ \epsilon(g)+ 2{\rm dim.} H_C^{2,0}(M,{\bf J}) + {\rm dim.} S^1.
\\ \nonumber
&\not= &{\rm dim.}~ \epsilon(g)+ {\rm dim.} S^2 + {\rm dim.} S^1.
\end{eqnarray}
\par
We mention   the   Teichm\"uller space $N$
on $K3$-surface \cite{besse}or $T^4$ \cite{myers},  which is defined by
  the quotient of K\"ahler-Einstein metrics of volume one modulo
the   diffeomorphisms which   induce the identity on the cohomology
group $H^2(M,Z)$ ;
\begin{equation}
 K3 : N \cong  SO(3,19) / SO(3) \times SO(19),
{}~
T^4 : N \cong  SO(4) \setminus GL(4)/ PSL(4,Z).
\end{equation}
\par The relations between the moduli spaces of
 $K(g)$   and $\epsilon(g)$  on $K3$ was already
derived\cite{besse}.
They  are some  manifolds
ewith holes or singularities\cite{kob,besse}.
There is a naturl compactification   of marked $\epsilon (g)$ on $K3$
by the method of Satake, Baily-Borel and  Mumford\cite{kob}.
These mathematical  results  will be important  for the further investigation
  into   ${\cal M}(\Sigma)$ and the evaluation of some observables.
\\
\par
Acknowledgment
\\
 We are grateful to  T. Ueno, A. Nakamichi,
Q-Han Park, S. Morita and  N. Sakai
for useful discussions.
We  thank  A. Futaki  most for pointing out
   the difference between $K(g)$  and ${\rm M} (\Sigma)$ and the
necessity of  an  almost complex structure with  vanishing real
first Chern class for the $\Lambda =0$ case.
\newpage
\begin{center}
{\small
\begin{tabular}{|c|c|c|c|}  \hline \hline
\multicolumn{4}{|c|}{ Table 1. A Dimension-Counting
of Fundamental Variables in Witten-type model
} \\ \hline \hline
\it $ \Sigma^k $ ~~~~~~~~~~~~~~~ ($3 \times 6= 18 $ )
       & $ \Psi^k $~~~~~~~~~~~~~~~ ($3 \times 6= 18$)  \\ \hline \hline
\it  {\rm diffeo.~ gauge~fix.~condi.}
   & red.~diffeo. gauge~fix.~condi.
\\
\it $D_0^\ast \Sigma^k = 0 ~~~~~~~~~~~~~~~~~~~~~~~~~~~~~(4)$
& $D_0^\ast \Psi^k = 0 ~~~~~~~~~~~~~~~~~~~~~~~~~~~~~(4) $
\\ \hline
\it  {\rm super. / red.~  gauge~ fix.~ condi. }
    & {\rm  eqs. of motion}
 \\
\it $D_1\Sigma^k=0 :~
 \{ d\Sigma^k=0 \}/
     \{ d^2 \Sigma^k=0  \} ~ (9)$
    &  $D_1\Psi^k=0 :~  \{ d\Psi^k=0 \}/
     \{ d^2 \Psi^k=0 \}  ~  (9)$
 \\
\it ~~~~~~~~~~~~~~~~~ ${}^{\rm t.f.} \Sigma^i \wedge \Sigma^j=0
{}~~~~~~~~~~(5) $
    & ~~~~~~~~~~~~~~~~~~~${}^{\rm t.f.} \Sigma^i \wedge \Psi^j=0
{}~~~~~~~~~~~(5) $
 \\ \hline \hline
\end{tabular}
}
\end{center}
\newpage
\begin{center}
\begin{tabular}{|c|c|c|c|c|}  \hline \hline
\multicolumn{5}{|c|}{Table 2. characteristic  number } \\ \hline \hline
\it {} &$T^4$ & K3 & $K3_{Z_2}$ & $K3_{Z_2 \otimes Z_2}$
\\ \hline
\it
 $\chi=c_2 $ &0&24&12&6
\\ \hline
\it
 $\tau$ & 0&-16&-8&-4
\\ \hline
\it
 $c_1$ & 0 & 0 &0 & 0
\\ \hline
\it
 $b_0(K^\ast)$ &1&1&0&$ -{1\over 2}$
\\ \hline
\it
 $b_1(K^\ast)$ & 4&0&0&0
\\ \hline
\it
 $b_2(K^\ast)$ & 6 &22& 12& 7
\\ \hline
\it
 $b_3(K^\ast)$ & 4 &0 &  0 &0
\\ \hline
\it
 $b_4(K^\ast)$&1&1&0&$-{1\over 2}$
\\ \hline
\it
 $h^0(M, \Theta)$ &  2&0&0&0
\\ \hline
\it
 $h^1(M, \Theta)$ &4&20&10&5
\\ \hline
\it
 $h^2(M, \Theta)$&2&0&0&0
\\ \hline
\it
 $b_0 $&1&1&1&1
\\ \hline
\it
 $b_1 $&4&0&0&0
\\ \hline
\it
 $b_2 $ & 6&22&10&4
\\ \hline
\it
 $b_{1,1}$ & 4&20&10&5
\\ \hline
\it
 $b_{2,0}$ & 1&1&0&$-1 \over 2$
\\ \hline
\it
  {\rm dim.} Isometry &4&0&0&0     \\ \hline\hline
\end{tabular}
\end{center}
\newpage

\end{document}